\documentclass{elsart5p}
\usepackage{amsmath,epsfig,amsfonts,amssymb}
\begin{document}

\begin{frontmatter}
\title{3D Building Model Fitting Using A New Kinetic Framework}

\author{Mathieu Br\'edif $^{a,b}$, Didier Boldo $^a$, Marc Pierrot-Deseilligny $^a$, Henri Ma\^itre $^b$}
\address{$^a$ IGN (French Mapping Agency), MATIS, 2/4 avenue Pasteur, 94165 Saint-Mand\'e, Cedex, France\\
      $^b$ Institut TELECOM - TELECOM ParisTech, LTCI UMR 5141 - CNRS, 46 rue Barrault, 75013 Paris, France}

\begin{abstract}
We describe a new approach to fit the polyhedron describing a 3D building model to the point cloud of a Digital Elevation Model (DEM). We introduce a new kinetic framework that hides to its user the combinatorial complexity of determining or maintaining the polyhedron topology, allowing the design of a simple variational optimization.

This new kinetic framework allows the manipulation of a bounded polyhedron with simple faces by specifying the target plane equations of each of its faces. It proceeds by evolving continuously from the polyhedron defined by its initial topology and its initial plane equations to a polyhedron that is as topologically close as possible to the initial polyhedron but with the new plane equations. This kinetic framework handles internally the necessary topological changes that may be required to keep the faces simple and the polyhedron bounded. For each intermediate configurations where the polyhedron looses the simplicity of its faces or its boundedness, the simplest topological modification that is able to reestablish the simplicity and the boundedness is performed.

\end{abstract}

\end{frontmatter}

\section{Introduction}

To satisfy the growing demand for 3D city models of increasingly better accuracy and higher level of detail, there is a solid research effort to model the real world using lidar data or satellite, aerial or ground imagery. To drive costs down, the algorithm must be automatic and make use of the coarser or inaccurate 3D city models that may already exist.

This fitting seems compulsory for state of the art building reconstruction methods\cite{Baillard,scholzevangool,tail} that proceed by first detecting a static set of geometric features like planes, points or lines and then determine the building topology using a combinatorial exploration. As this combinatorial exploration time explodes when the number of initial geometric features increases, the number of such detected features is kept low. Together with the fact that the geometry of the detected features is static, this small number of possibly inaccurate geometric features yield approximate reconstructions. Fitting these approximate geometries to the available data as a post process will improve the geometric accuracy of the reconstructed models. This paper presents a fully automatic method to fit an existing inaccurate building model to a real world data set. 
Previous works\cite{TV05} mainly focus on optimizing the point coordinates of the model and do not question the topology of the 3D model. They keep the topology fixed and either fail or stop before convergence when it occurs that the initial topology was wrong as illustrated in Figure~\ref{mb3_hardedgecollapse}.

\begin{figure}
 \centering\includegraphics[width=4cm]{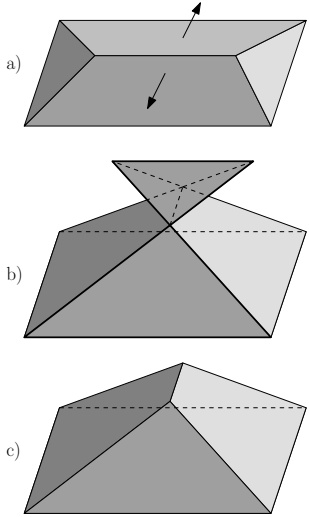}
\caption{a) A polyhedron with two faces that are evolving with an increasing slope. b) The resulting polyhedron if no topological modification is performed: faces are no longer simple and an inside-out tetrahedron is present. c) At the time in the evolution when the problematic edge length was null, a topological flip of this edge has been performed to keep the faces simple just after the singularity.\label{hardedgecollapse_inversion}}
\end{figure}

\begin{figure}
\centering\includegraphics[width=4cm]{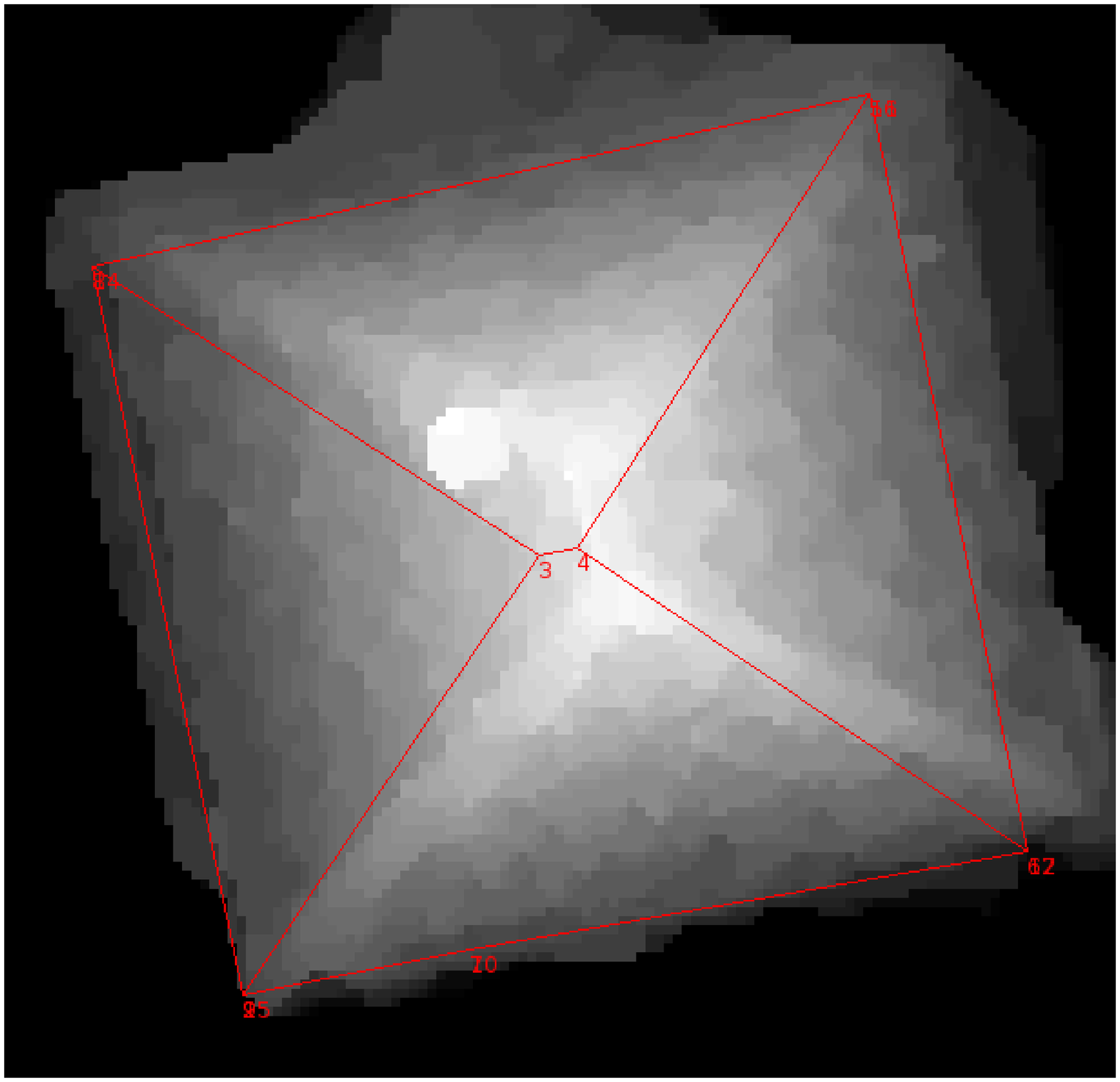}\centering\includegraphics[width=4cm]{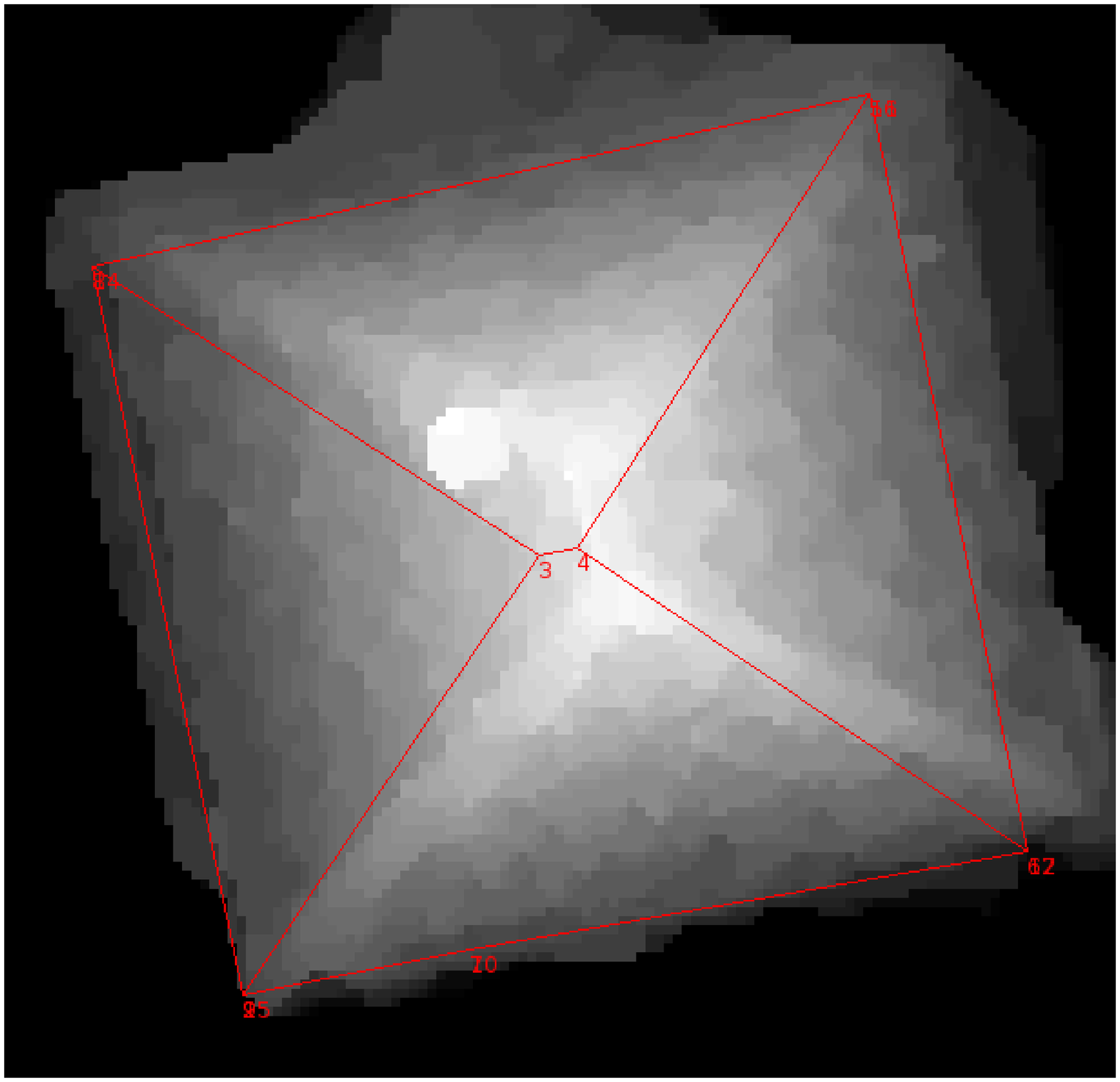}
 \caption{A simple applicative instance of the problem described in Figure~\ref{hardedgecollapse_inversion}: this figure shows the building wireframe superimposed over the fitted DEM. The left image shows the initial building wireframe with a topologically erroneous roof top edge. The right image is the result of our fitting process: the roof top edge has been topologically flipped and the other roof edges are no longer required to go through the vertices of the roof plan, allowing a better fit.\label{mb3_hardedgecollapse}}
\end{figure}

To achieve 3D building model fitting without having to keep the topology fixed, this paper introduces a new generic framework to fit a polyhedron model to data. It makes use of an initial polyhedron and a mapping from the initial supporting planes of its faces to a set of target supporting planes. It continuously interpolates the representation from the initial polyhedron to a novel polyhedron built on the target plane equations, with minimal topological modifications. During this continuous evolution, the algorithm is able to handle implicitly the topological changes that are required to keep a 3D model well formed when the geometry of the 3D model is altered, as illustrated in figure~\ref{hardedgecollapse_inversion}. This makes the design of a variational shape optimization possible. To our knowledge, only the variational shape approximation approach\cite{vsa} achieves the same goal but it only maintains a partition of the fitted data. The resulting polyhedron is only exported as a post process. Furthermore, it cannot guarantee that a partition cell is exported as a single face or even that the exported faces will be supported by a common plane. By contrast, our approach maintains a polyhedron throughout the optimization, thus avoiding the final export. Besides, the guarantee of our framework that data that has been fitted to a plane will be represented by faces supported by that plane is particularly relevant when there is some semantics attached to the faces of the polyhedron: a roof plane will be modeled with a single face and not a set of nearly coplanar faces.

The next section describes the new kinetic framework. Then, an application of this framework to fit 3D building models is presented and discussed.

\section{Kinetic Framework}

For a better understanding of the framework, let us first introduce the key definitions about polyhedra that will be used to represent the 3D models.

\subsection{Definitions}

A \textbf{polyhedron} is defined as an oriented piecewise-planar 2-manifold without boundary. Its description is two-fold: its \textbf{topology} is given by the incidence graph linking its combinatorial elements (vertices, edges and faces), and its \textbf{geometry} can be given by the 3D coordinates of all its vertices (its point cloud).

Note that the \textbf{dual geometry} may be given instead: \textit{i.e.} the equations of the planes that support the faces of the polyhedron. Instead of the 3D point cloud of the \textbf{primal geometry}, this defines a 3D plane arrangement. The topology then only states which faces of the plane arrangement are part of a polyhedron face, or which 3D cell of the arrangement are inside the polyhedron.

We define the following \textbf{topological} properties on polyhedron elements:
\begin{itemize}
 \item A \textbf{triplanar vertex} is adjacent to exactly 3 faces. It thus maps to a triangular face in the dual polyhedron. In the general case, only triplanar vertex coordinates may be uniquely determined from the equations of their adjacent faces.
\item A \textbf{triangulated face} is a face that is subdivided by recursively splitting it with edges between 2 of its vertices until the face made of triangles only. Those subdivision edges are called \textbf{soft} edges as opposed to the initial non-subdividing \textbf{hard} edges of the polyhedron. Note that this is only the definition of a topological triangulation: there is no geometric requirement on the triangle orientations.
 \item An edge of a polyhedron may be defined by its 2 vertices or its 2 incident faces. A \textbf{non-degenerate edge} is an edge for which the only faces that are adjacent to both vertices of the edge are its incident faces. The topology of a polyhedron can always be changed to describe the same surface without degenerate edges, yielding a simpler representation.
\end{itemize}

The following properties only involve \textbf{geometric} requirements:
\begin{itemize}
 \item A \textbf{bounded vertex} lies in the interior of a given bounding convex polyhedron $V^\infty$, like the bounding box $V^\infty=[-M,M]^3$.
 \item A \textbf{simple face} is a face of a polyhedron, the geometry of which is a simple (non auto-crossing) 3D polygon. A face is simple if and only if it can be triangulated so that its 3D triangles have an orientation consistent with the orientation of the polyhedron. 
\end{itemize}
A \textbf{triplanar, bounded, simple, triangulated or non-degenerate polyhedron} is defined by extension when all its qualified elements (vertex, edge or face) verify the property.

\subsection{Problem Statement}
The task that we are willing to perform is the following: given a simple bounded polyhedron $P_0$ and a target plane equation for each of its faces (an updated dual geometry), output a simple bounded polyhedron $P_1$ which match the target dual geometry (supporting planes are the given target planes) and the topology of which is as close as possible to the initial topology.

This problem can be restated as a 0-1 coloring of the cells of the plane arrangement of the target planes and the planes of $V^\infty$. From such a coloring, the 0.5 level set is a set of polyhedra whose faces are supported by the target planes. The boundedness can be translated as a 0-coloring of the unbounded 3D cells of the arrangement. A straightforward 0-1 coloring is the volumetric thresholding: for each cell $C$, the ratio of volume of the intersection $P_0\cap C$ of the initial polyhedron $P_0$ to the volume of the cell $C$ is computed. Then a predefined threshold could assign a value of 0 or 1 to the cell.

However, it could result in overly complex shapes due to the oversegmentation of the volume $V^\infty$ given by the plane arrangement. Worse, the initial topology is not directly taken into account, leading to counter intuitive results. Furthermore, the simplicity can be enforced by splitting the polyhedra at non-manifold vertices or edges, but the result will be a set of polyhedra and not a single polyhedron.

\subsection{Kinetic Data Structure}
The generic \textit{kinetic framework} that has been developed in \cite{basch97,basch99} appears to be a good solution for our problem. The idea is to move continuously the geometry of the data from the initial geometry at time $t=0$ to the target geometry at $t=1$. In a kinetic framework, a global geometric property is maintained by constructing and maintaining a kinetic data structure (KDS) throughout the time evolution. The purpose of the KDS is to track the list of geometric tests or predicates that are required to prove the global property. This data structure is responsible for ensuring that this finite set of local tests or predicates, called \textit{certificates}, which together prove the global geometric properties, remain valid. These certificates are typically polynomial or rational functions in terms of the interpolation time $t$, that are respectively valid, degenerate or invalid when their signs are respectively positive, null or negative. Roots (and poles) of these certificates are called failing times, because the continuous nature of the certificates ensure that the sign of a certificate remains constant between failing times. During the simulation, the time does not evolve continuously: the KDS computes the failing times of each of its certificates and orders them in a priority queue. The interpolation time $t$ is then iteratively advanced to the closest failure time in the future. At that time, a certificate fails and the KDS has to make minimal changes to the topology of the object and has to update its internal proof accordingly. These updates reestablish a set of valid certificates, so that time can then be advanced either to the next failing time of one of the certificates or to the evolution ending time $t=1$.

We adapted this generic framework provided by the kinetic data structure package of the CGAL library\cite{cgal:f-kdsf-07} to our problem, the interpolation between the initial and target dual geometry is introduced in the next subsection, using homogeneous coordinates.

\subsection{Oriented Projective Geometry}

The geometry of Cartesian three-space $\mathbb R^3$ is greatly simplified by using the oriented projective space $\mathbb T ^3$. An in-depth presentation of this space  $\mathbb T ^3$ can be found in \cite{orientedprojectivegeometry}. It retains the powerful representation and unification properties of the unoriented projective space $\mathbb P ^3$ while preserving the orientation and separability of the Cartesian space $\mathbb R^3$.

The homogeneous coordinates $\vec{P}=[x,y,z,w]=[\vec{p},w]$ refer, if $w\neq 0$, to the 3D point of Cartesian coordinates $\frac{\vec{p}}{w}=(\frac{x}{w},\frac{y}{w},\frac{z}{w})$. The multiplication of the homogeneous coordinates $\vec{P}$ by any non-zero factor leaves the Cartesian 3D point unchanged.

A plane equation is given by the homogeneous coordinates $\vec{N}=[a,b,c,d]=[\vec{n},d]$. It refers to the planar set of points $\vec{P}$ that verify $\vec{P}\cdot\vec{N}=0$, which is equivalent, if $w\neq0$, to the Cartesian equation $\frac{\vec{p}}{w}\cdot\vec{n}+d=0$.
The normal of the plane is thus encoded in $\vec{n}=(a,b,c)$ and, if $\vec{n}$ is normalized, $d$ represents the signed distance between the plane and the origin. The multiplication of the homogeneous coordinates $\vec{N}$ by any non-zero factor leaves also the plane unchanged.
The direction of the normal vector $\vec{n}$ induces an orientation of the plane $\vec{N}$: a plane $\vec{N}$ multiplied by a negative factor thus refers to the same plane but with a reversed orientation.

The signed distance from a point to a plane is computed with $\frac{\vec{P}}{w}\cdot\frac{\vec{N}}{\|\vec{n}\|}$. It follows that to test on which side of a plane a point is, one has to evaluate:

\begin{eqnarray} side&=&sign\left(\frac{\vec{P}}{w}\cdot\vec{N}\right)
\end{eqnarray}
A zero means that the points in on the plane, a positive or negative sign denotes respectively a point in the halfspace targeted by $\vec n$ or the other halfspace.

The orientation of triangle $\vec{P}_0\vec{P}_1\vec{P}_2$ is determined by the sign of the following 4 by 4 determinant where $\vec{P}_{\vec{n}}=[\vec{n},0]$ is the point at infinity along the direction $\vec{n}$ and the $\frac{\vec{P}_i}{w_i}=[\frac{x_i}{w_i},\frac{y_i}{w_i},\frac{z_i}{w_i},1]$ are the normalized point coordinates:

\begin{eqnarray} orientation&=&sign\left(\begin{vmatrix} \vec{P}_{\vec{n}}&\frac{\vec{P}_0}{w_0}&\frac{\vec{P}_1}{w_1}&\frac{\vec{P}_2}{w_2}\end{vmatrix}\right)
\end{eqnarray}

 A positive, null or negative  determinant denotes respectively a direct, aligned or indirect triangle with respect to the orientation defined by $\vec{n}$.

The linear combination $\vec{P}_t=(1-t).\vec{P}_0+t.\vec{P}_1$ with $t\in[0,1]$ spans the segment from $\vec{P}_0$ to $\vec{P}_1$. Likewise, the linear combination $\vec{N}_t=(1-t).\vec{N}_0+t.\vec{N}_1$ with $t\in[0,1]$ has a geometric interpretation too: if the planes are not parallel, they intersect at a common line $\vec{L}$ and the linear combination lets the plane $\vec{N}_t$ rotate around the line $\vec{L}$ from $\vec{N}_0$ to $\vec{N}_1$.
If they are parallel ($\vec{n}_0$ and $\vec{n}_1$ are collinear), the linear combination spans the set of planes that translate from $\vec{N}_0$ to $\vec{N}_1$.  A special case that will be avoided in our framework is to compute the linear combination of 2 planes that are parallel with opposite orientations because, this is the only case where there is a $t\in[0,1]$ for which $\vec{n}_t=(1-t).\vec{n}_0+t.\vec{n}_1=\vec 0$ which is the plane containing all the points at infinity.

The homogeneous coordinates $\vec P= [x,y,z,w]$ of a triplanar vertex can be computed from the plane equations $\vec N_0$, $\vec N_1$, $\vec N_2$ of its 3 adjacent faces. They are cofactors $cof_{1,j}$ of the first column of the 4 by 4 matrix $\begin{bmatrix}\vec0&\vec N_0&\vec N_1&\vec N_2\end{bmatrix}$:
\begin{eqnarray}
 \vec{P}=[x,y,z,w]&=&[cof_{1,1},cof_{1,2},cof_{1,3},cof_{1,4}]\\
\textrm{For instance, }w&=&cof_{1,4}=-\begin{vmatrix}\vec n_0&\vec n_1&\vec n_2\end{vmatrix}
\end{eqnarray}

It follows that the Cartesian coordinates $\frac{\vec p}{w}$ of a point are rational functions in terms of the 12 coefficients of the 3 adjacent planes. A triplanar point is thus defined everywhere except at the poles of $\frac{\vec p}{w}$ where it goes to infinity. A point is however undefined if its 4 polynomial coordinates in terms of the adjacent plane coefficients are null. This can be avoided if either the initial or the target geometry has no subset of 3 planes that have linearly dependant normals.

From this formulation, it appears that the conservative restriction on the target dual geometry is to disallow orientation reversals $\vec n\rightarrow\lambda\vec n$ with negative or null $\lambda$ and degenerate dual geometries where there are at least 3 planes with linearly dependant normals. A less conservative solution is to first evolve to an intermediate non degenerate dual geometry in which planes with orientation reversals are slightly rotated, and then evolve to the target dual geometry.

\subsection{Polyhedron Triplanarization}
\begin{figure}
 \centering\includegraphics[width=6cm]{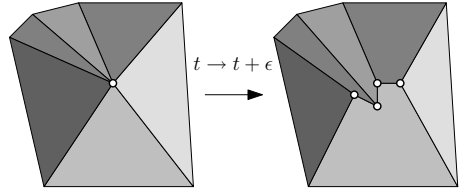}
\caption{Triplanarization of a 6-planar vertex.\label{triplanar}}
\end{figure}

\begin{figure}
 \centering\includegraphics[width=6cm]{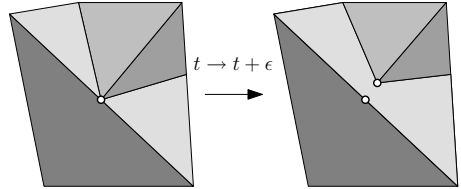}
\caption{Biplanar vertex resulting from a triplanarization. The biplanar vertex is removed at a post-processing step.\label{biplanar}}
\end{figure}

To be able to define the point coordinates from the adjacent planes uniquely, the polyhedron must be modified to be triplanar. Because the polyhedron is closed, any vertex with less than 3 different adjacent planes can safely be removed without altering the described manifold: isolated points (arity 0) and point in the interior of a face (arity 1) are simply removed. Points lying on the interior of an edge (arity 2) are also removed by collapsing one of the two adjacent edges. Vertices that are adjacent to more than 3 planes are splitted into triplanar (Fig.~\ref{triplanar}) or biplanar (Fig.~\ref{biplanar}) vertices. This splitting introduces zero length edges if the planes are concurrent (intersecting at a single point).

However, a point may be overdefined by planes that are nearly but not exactly concurrent because of arithmetic errors. Worse, it may be in the degenerate case where the planes are instantaneously concurrent but at a time immediately after in the evolution, planes will no longer be concurrent and thus the point will no longer be properly defined. In this case, the exact (\textit{e.g.} with infinite precision arithmetics) geometric predicates provided by the CGAL library\cite{cgal} allow to choose the splitting into triplanar or biplanar vertices that will yield simple faces. To be precise, this splitting problem is a subproblem of the 0-1 coloring of the plane arrangement of all the planes that considers only the adjacent planes and for which the unbounded cell coloring is given by their position relative to the adjacent oriented planes.

\begin{figure}
 \centering\includegraphics[width=6cm]{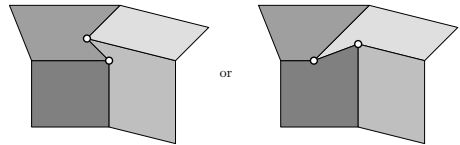}
\caption{This is the only case (up to symmetries) were a quadriplanar vertex admits 2 distinct triplanarizations.\label{ambiguity}}
\end{figure}

This problem is solved in our algorithm by a dual \textit{ear-cutting} like approach. A triangulation can be performed by successively and iteratively cutting ears (direct triangles with 2 consecutive edges of the polygon) until the polygon is a triangle. Our dual problem of triplanarizing a vertex is performed likewise by successively and iteratively joining together 2 consecutive edges of the overdefined vertex to produce a new vertex. This vertex may be triplanar (Fig.~\ref{triplanar}) or biplanar when an adjacent plane is placed between 2 faces supported by the same plane (Fig.~\ref{biplanar}). Instead of the orientation requirement of the triangulation problem, the triplanarization requires that the resulting adjacent faces are simple. Like the triangulation, there are cases were multiple triplanarizations satisfy the requirement (Fig.~\ref{ambiguity}). Our algorithm does not currently take this possible multiplicity into account and applies the first triplanarization found that produces simple adjacent facets. For extremal vertices, for which the polyhedron lies localy only on one side of a tangent plane, the triplanarization amounts to computing a straight skeleton\cite{eppstein99raising,straightskeleton} weighted in function of the dihedral angles of these faces with such a tangent plane.

\subsection{Algorithm Overview}

The naive and error prone approach of adding constant or even adaptative time steps, then verifying the simplicity and boundedness lacks the knowledge of what exactly went wrong since the last iteration, whereas this kinetic approach is able to handle well identified problems one at a time. The algorithm follows this procedure:
\begin{itemize}
 \item Vertices that are not triplanar are triplanarized to ensure that the point coordinates can be computed from the moving plane equations.
 \item Faces that are not triangular are triangulated to provide the proof of simplicity.
 \item Certificates that have been invalidated by topological changes are updated and their next failing times are computed, to update the proofs of simplicity and boundedness.
 \item Time is stepped directly to the next failing time $t$ or the algorithm stops if $t\geq1$ or if there is no failing time in the future.
 \item Until the next failing time is not immediate, the necessary topological updates are performed.
\end{itemize}

To avoid infinite loops where topological changes create recursively immediately failing configurations, the triplanarization and triangulations require the triangles to be direct and does not allow degenerate aligned triangles. For this to be possible, there must be no certificate of which the current time is a root. So all the simultaneous current events are processed until the next failing time is not immediate, before looping back to the triplanarization and the triangulation. A side effect of this batch processing of the simultaneous events is to avoid the formation of degenerate edges.

\subsection{Boundedness Certificates}
A polyhedron is bounded if all of its vertices are bounded. The convex bounding volume $V^\infty$ is given by a set of oriented plane equations $\vec N_i^\infty$ that point toward the exterior of the bounded volume. A vertex $\vec P$ remains inside $V^\infty$ if and only if the \textit{side} certificate is negative for all faces of the bounding volume: $\forall i,\frac{\vec P}{w}\cdot\vec N_i^\infty\leq0$. Thus for each vertex $\vec P$ of the polyhedron and each plane of the convex bounding volume $V^\infty$, we can define its boundedness certificate $-\frac{\vec P}{w}\cdot\vec N_i$. These certificates are not negative for each vertex and each bounding plane, if and only if the polyhedron is bounded. Since the plane equations are the result of a linear interpolation, their coefficients are polynoms of maximum degree 1. If the bounding volumes is kept constant, that means that the certificate is a rational function and the maximum degrees of its numerator and denominator are both 3. Computing the failing times amounts to finding roots of the numerator with odd multiplicity, while checking if they are also roots of the denominator to eventually decrease the multiplicity by the root multiplicity at the denominator. Sturm's theorem and interval arithmetics are used in the CGAL library\cite{cgal} to provide efficiency, infinite accuracy and lazy comparisons between the roots.

\subsection{Orientation Certificates}

The simplicity of the faces of the polyhedron is proven by their triangulations with triangles that are oriented consistently with the oriented supporting planes. The KDS will keep such a triangulation for each of the faces and compute for each of its triangles the \textit{orientation} certificate $\begin{vmatrix} \vec{P}_{\vec{n}}&\frac{\vec{P}_0}{w_0}&\frac{\vec{P}_1}{w_1}&\frac{\vec{P}_2}{w_2}\end{vmatrix}$ that tests whether the triangle $\vec{P}_0\vec{P}_1\vec{P}_2$ is direct, degenerate or indirect with respect to the supporting oriented plane $\vec{N}=[\vec{n},d]$. The coordinates of $\vec{P}_{\vec{n}}$ are polynoms of degree at most 1 and the 3 other columns are rational functions with numerators and denominators of degree at most 3. The reason why the vertex are kept bounded is to avoid the tricky situation where the Cartesian coordinates $\frac{\vec{P}}{w}$ of a point are evaluated at one of its poles. Keeping all the vertex bounded ensures that sign changes in the orientation certificates are only due to its roots and not its poles. This yields a rational function that has a numerator of degree at most 10 and a factorization of the denominator into 3 polynoms of degree at most 3. The sign of the certificate can be determined by the independent inspection of the signs of the factors of its numerator and denominator and the evaluation of the root multiplicities if the current time is a root of the denominator. This speeds up the evaluation compared to computing blindly roots of the multiplication of the numerator by the denominator because this polynom can be of degree 10+3+3+3=19 !
It is even possible to factorize the numerator, because there is a polynom $\lambda$ that links the estimated normal $\vec{n}_{\vec{P}_0\vec{P}_1\vec{P}_2}$ to the plane normal $\vec n$:
\begin{equation}
\vec{n}_{\vec{P}_0\vec{P}_1\vec{P}_2}=\left(\frac{\vec{p_1}}{w_1}-\frac{\vec{p_0}}{w_0}\right)\wedge\left(\frac{\vec{p_2}}{w_2}-\frac{\vec{p_0}}{w_0}\right)=\frac{\lambda.\vec{n}}{w_0.w_1.w_2}
\end{equation}
It follows that the numerator of the certificates can be factorized to prove that sign changes of the numerator are only due to sign changes of $\lambda$:
\begin{equation}
\begin{vmatrix} \vec{P}_{\vec{n}}&\frac{\vec{P}_0}{w_0}&\frac{\vec{P}_1}{w_1}&\frac{\vec{P}_2}{w_2}\end{vmatrix}=\vec{n}\cdot\vec{n}_{\vec{P}_0\vec{P}_1\vec{P}_2}=\frac{\lambda.\left(\vec{n}\cdot\vec{n}\right)}{w_0.w_1.w_2}
\end{equation}

\begin{figure}
 \centering\includegraphics[width=3cm]{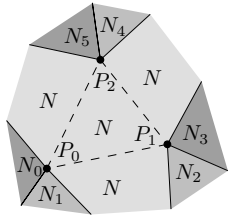}
 \caption{Triangle with 3 soft edges and without hard edges.\label{0hard}}
\end{figure}

If all the 3 edges of the triangle are soft (triangulation) edges and not hard edges of the polyhedron (Fig.~\ref{0hard}), the $\lambda$ polynom of the numerator of the certificate cannot generally be further factorized. An efficient way to compute $\lambda$ in the general case is yet to be found. In the meantime, roots are searched for the unfactored polynom $\begin{vmatrix} \vec{P}_{\vec{n}}&\vec{P}_0&\vec{P}_1&\vec{P}_2\end{vmatrix}$.

\begin{figure}
 \centering\includegraphics[width=3cm]{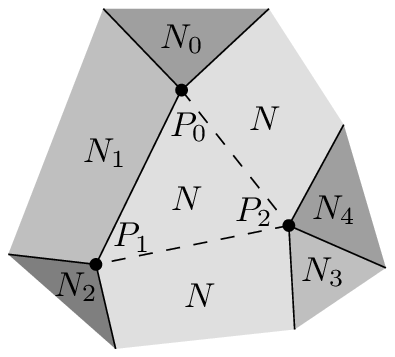}
 \caption{Triangle with 2 soft edges and 1 hard edge.\label{1hard}}
\end{figure}

\begin{figure}
 \centering\includegraphics[width=3cm]{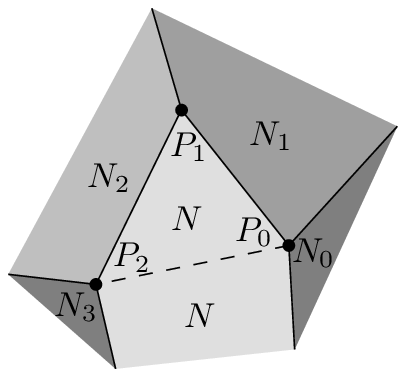}
 \caption{Triangle with 1 soft edge and 2 hard edges.\label{2hard}}
\end{figure}

When there is a mix of hard and soft edges (Fig.~\ref{1hard}~and~\ref{2hard}), $\lambda$ can be further factorized into 2 polynoms of degree at most 4, improving the root finding time:
\begin{eqnarray}
\lambda_{Fig.~\ref{1hard}}&=&-\begin{vmatrix}\vec{N}&\vec{N_0}&\vec{N_1}&\vec{N_2}\end{vmatrix}.\begin{vmatrix}\vec{N}&\vec{N_1}&\vec{N_3}&\vec{N_4}\end{vmatrix}\\
\lambda_{Fig.~\ref{2hard}}&=&-\begin{vmatrix}\vec{N}&\vec{N_0}&\vec{N_1}&\vec{N_2}\end{vmatrix}.\begin{vmatrix}\vec{N}&\vec{N_1}&\vec{N_2}&\vec{N_3}\end{vmatrix}
\end{eqnarray}

\begin{figure}
 \centering\includegraphics[width=3cm]{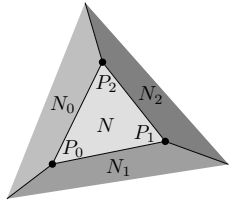}
 \caption{Triangular face without soft edges and with 3 hard edges.\label{3hard}}
\end{figure}

Finally, if all the edges are hard (Fig.~\ref{3hard}), then $\lambda_{Fig.~\ref{3hard}}=-\begin{vmatrix}\vec{N}&\vec{N_0}&\vec{N_1}&\vec{N_2}\end{vmatrix}^2$. Being the opposite of a square, $\lambda_{Fig.~\ref{3hard}}$ will not provoke any sign change, so there is no point in computing its roots.


\subsection{Boundedness Event Processing}
The basic idea is to cut the failing vertex by the bounding plane. However, the application that has driven our research cannot produce points that go to infinity, so it has not been implemented yet.

\subsection{Orientation Event Processing}
Since the polyhedron remains bounded, the 3 vertices of the failing triangle have finite coordinates. The failing of the triangle certificate means that those 3 points are aligned. The topological update is simply a topological translation of the geometric singularity.

\begin{figure}
 \centering\includegraphics[width=7cm]{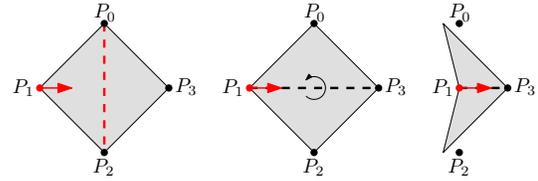}
 \caption{Example evolution of a vertex colliding with the opposite soft edge\label{softedgecollide}}
\end{figure}

\begin{figure}
 \centering\includegraphics[width=6cm]{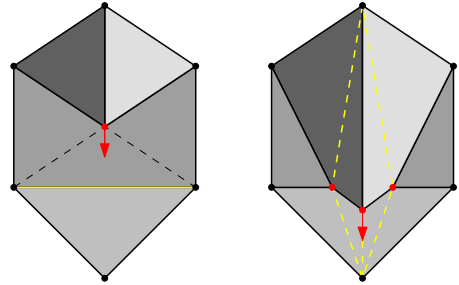}
\caption{Example evolution of a vertex colliding with the opposite hard edge\label{hardedgecollide}}
\end{figure}

In general, the 3 points are distinct, meaning that a vertex is colliding with the opposite edge. The polyhedron is updated by affecting the failing triangle to the plane supported by the facet that is opposite to the colliding vertex and by flipping the colliding edge. Example results at $t+\epsilon$ is sketched with a hard edge (Fig.~\ref{hardedgecollide}) and a soft edge (Fig.~\ref{softedgecollide}).

\begin{figure}
 \centering\includegraphics[width=7cm]{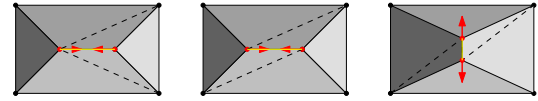}
\caption{Example evolution of a collapsing hard edge\label{hardedgecollapse}}
\end{figure}

In the particular case where 2 or 3 points are equal, either an edge or the entire face is collapsing.
 The polyhedron is updated by topologically collapsing one of the failing edges. An example result at $t+\epsilon$ is sketched in Fig.~\ref{hardedgecollapse}.

\subsection{Complexity}
Guibas and Bash introduced in \cite{basch97,basch99} a terminology to qualify the complexity of a kinetic data structure:
\begin{itemize}
 \item Our data structure is \textbf{compact} (linear in size), since a triangulation is linear in the size of its polygon. The maximum number of events in the queue is one orientation event per triangle and one above event per vertex and per bounding plane.
 \item An external event is an event that is required to update the topology to maintain the desired property, whereas an internal event is only needed by the KDS to update its proof (the triangulation in our case) but does not require any modification of the maintained object (the polyhedron here). The \textbf{efficiency} measures the ratio of external events to total events (internal and external). The only internal event is the soft edge collision.
 \item The \textbf{responsiveness} qualify the time complexity of processing a certificate failure. Processing the orientation events takes constant time, and processing the boundedness event takes time proportional to the degree of the vertex in the triangulated polyhedron. However, the triplanarization and triangulation steps may also take linear time.
 \item The KDS is as \textbf{local} as it could be: a number of events that depend on a single plane is linear relative to the complexity (the number of edges and vertices) of the faces of the polyhedron supported by this plane.
\end{itemize}

There are prospective alternatives to tackle this problem that can also detect general self intersections rather than only intersection of edges of a face, which is not possible in the current framework:
\begin{itemize}
 \item A kinetic 3D plane arrangement could keep track of the 0-1 coloring directly. However the number of its internal events may slow down drastically the algorithm.
 \item A kinetic tetrahedralization constrained by the polyhedron would be compact and relatively efficient but there is no guarantee that a polyhedron can be partitionned into tetrahedra without introducing utility vertices\cite{toussaint93tetrahedralization}. Dealing with this vertices may be tricky.
 \item An extension of our KDS to deal with self intersections between vertices that share at least 1 supporting adjacent plane seems possible. Instead of maintaining separate triangulations for each face, there would be for each plane a constrained triangulation of the convex hull of the faces supported by this plane. The triangulation would be constrained so that edges of the faces are also edges of the triangulation.
 \item Finally, a promising possibility is to schedule all the possible events (vertex-vertex, vertex-edge, vertex-face and edge-edge collisions) between all possible pairs without the help of a triangulation or a tetrahedralization. This would lead to an efficient framework without internal events, but it would no longer be compact (linear in size) : the number of events would be quadratic.
\end{itemize}

\section{Application: 3D Building Fitting}

To our knowledge, previous methods\cite{TV05} to fit a 3D building model to some data like calibrated images, or a laser point cloud were not questioning the topology of the initial model. They assumed that the ideally fitted model had the same topology as the given model. Therefore, our research efforts have been driven toward a framework that is able to infer implicitly the minimal topological changes from a geometric evolution.

\subsection{Optimization}

The proposed optimization is a local optimization. The algorithm is iterating optimization steps until either a maximum number of iterations are performed, or the computation time has been too long, or the quality of the fitting is not increasing sufficiently, meaning that the optimization has converged. Thus, there is no guarantee to obtain a global minimum, but a local minimum close to the approximate model appears to be a good guess in practice.

During an optimization step, the data is partitionned by the current 3D model: the DEM is partitionned by a vertical projection of the 3D model into regions that correspond to points of faces belonging to a common supporting plane. The plane equations $[a,b,c,d]$ are then estimated independently using a robust $\mathcal L_1$ estimator over the corresponding region of the DEM. A known limitation of this geometric optimization step is hence the impossibility to reconstruct vertical or bottom-facing roof planes. Although vertical faces of the polyhedron are not a special case of the new kinetic framework, it should be noted that the proposed geometric optimization is not able to move the vertical faces. A possible extension could be to no longer partition the data by a vertical projection but using a Vorono� decomposition the three-space, affecting the data to the closest polyhedron face in $\mathbb R ^3$. A simplifying by-product of the fixed vertical planes delimiting the floor plan and the top-facing orientation of the roof planes is that no vertex can go to infinity during the interpolation. The polyhedron will always remain bounded by any sufficiently large volume, so there is no point in scheduling and processing the boundedness events in a interpolation driven by this optimization. They are exposed in this paper for completeness.

The second step is obviously to use the kinetic polyhedron data structure described in this paper to drive the polyhedron from its current geometry and topology to the newly estimated geometry while minimizing the topological changes.

This algorithm is easily adaptable to lidar data, with a careful dealing of the possible planar inhomogeneity of the lidar points to prevent some estimation bias.

\subsection{Results}

\begin{figure}
\centering\includegraphics[width=4cm]{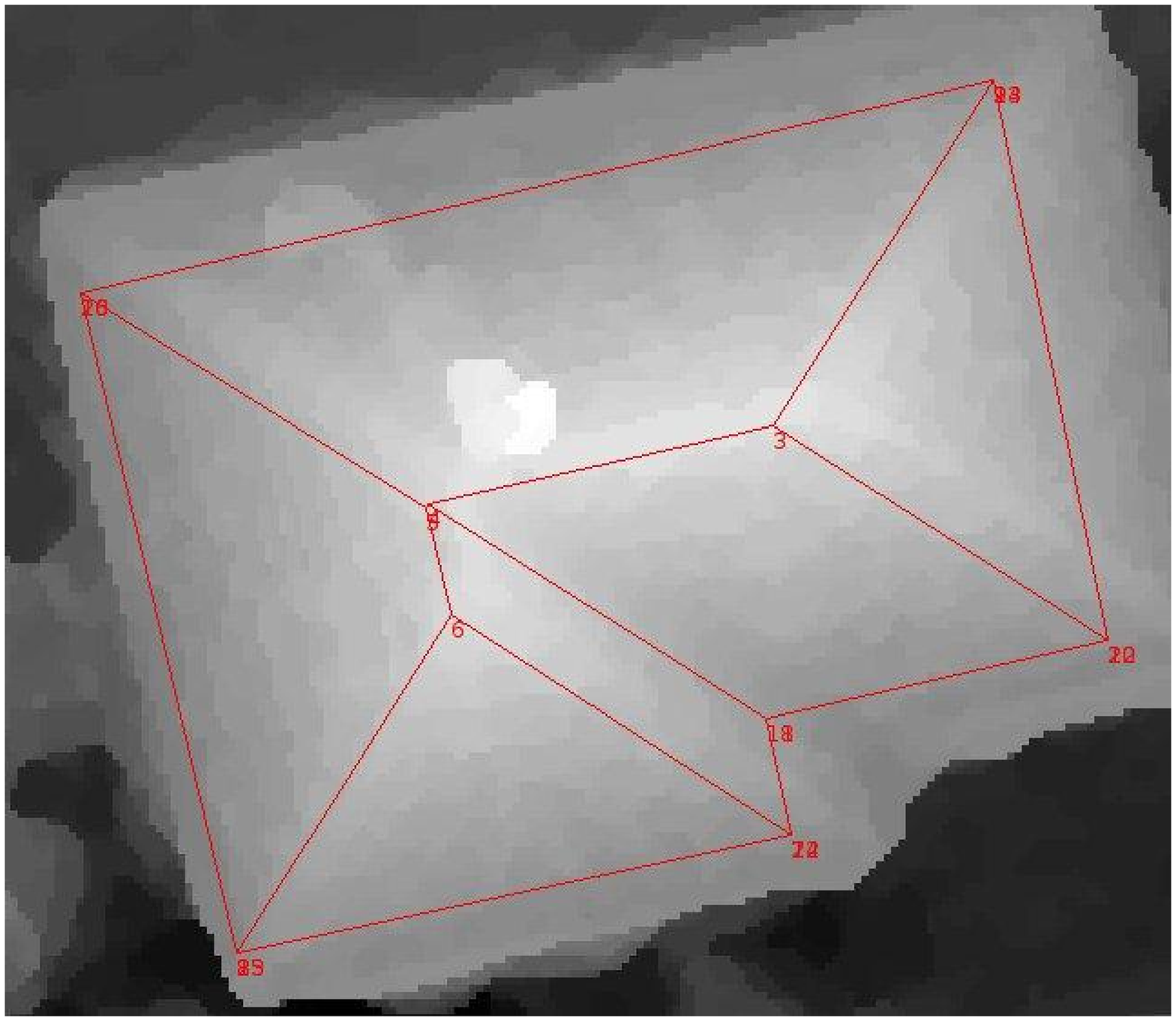}\centering\includegraphics[width=4cm]{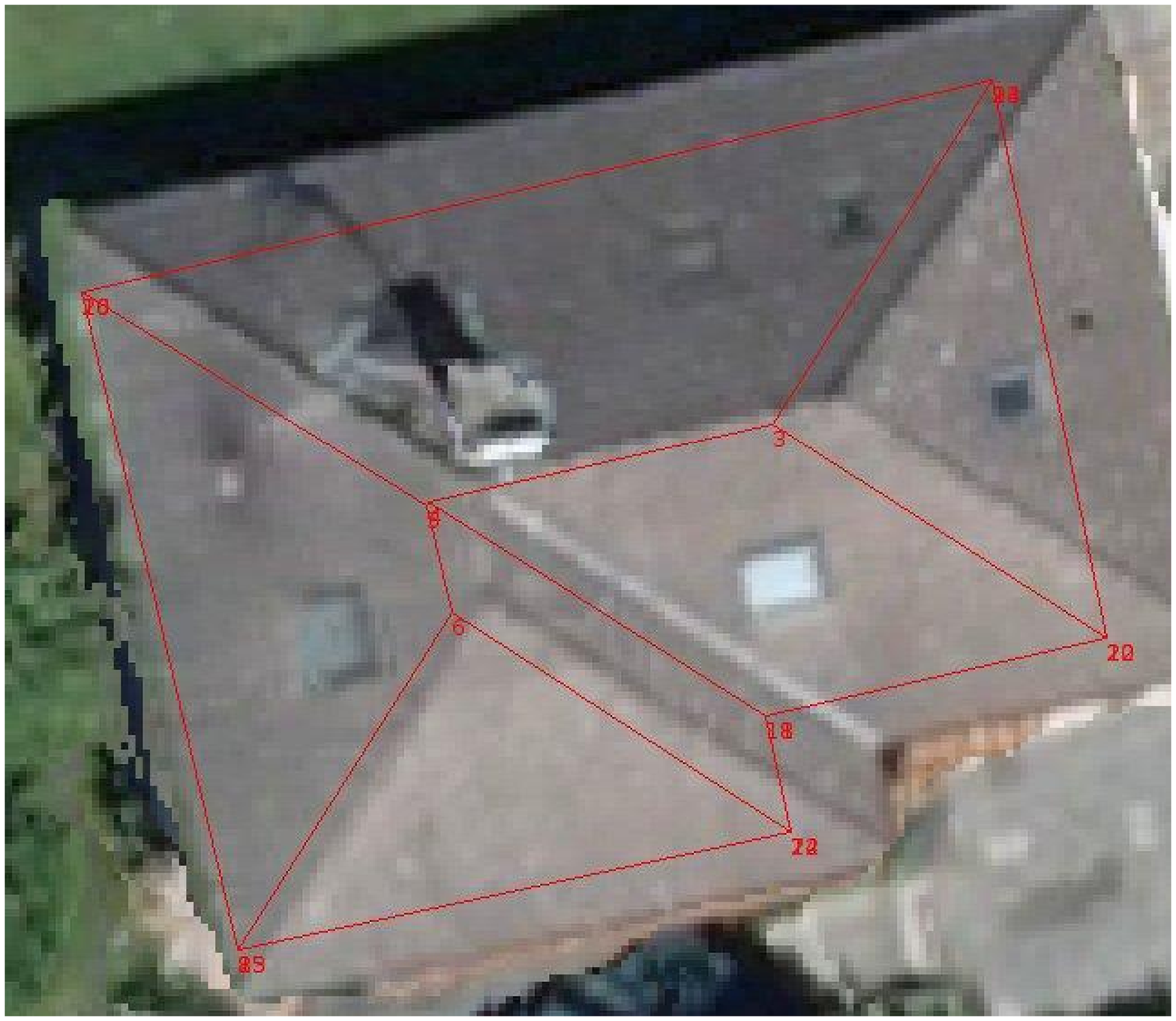}
\centering\includegraphics[width=4cm]{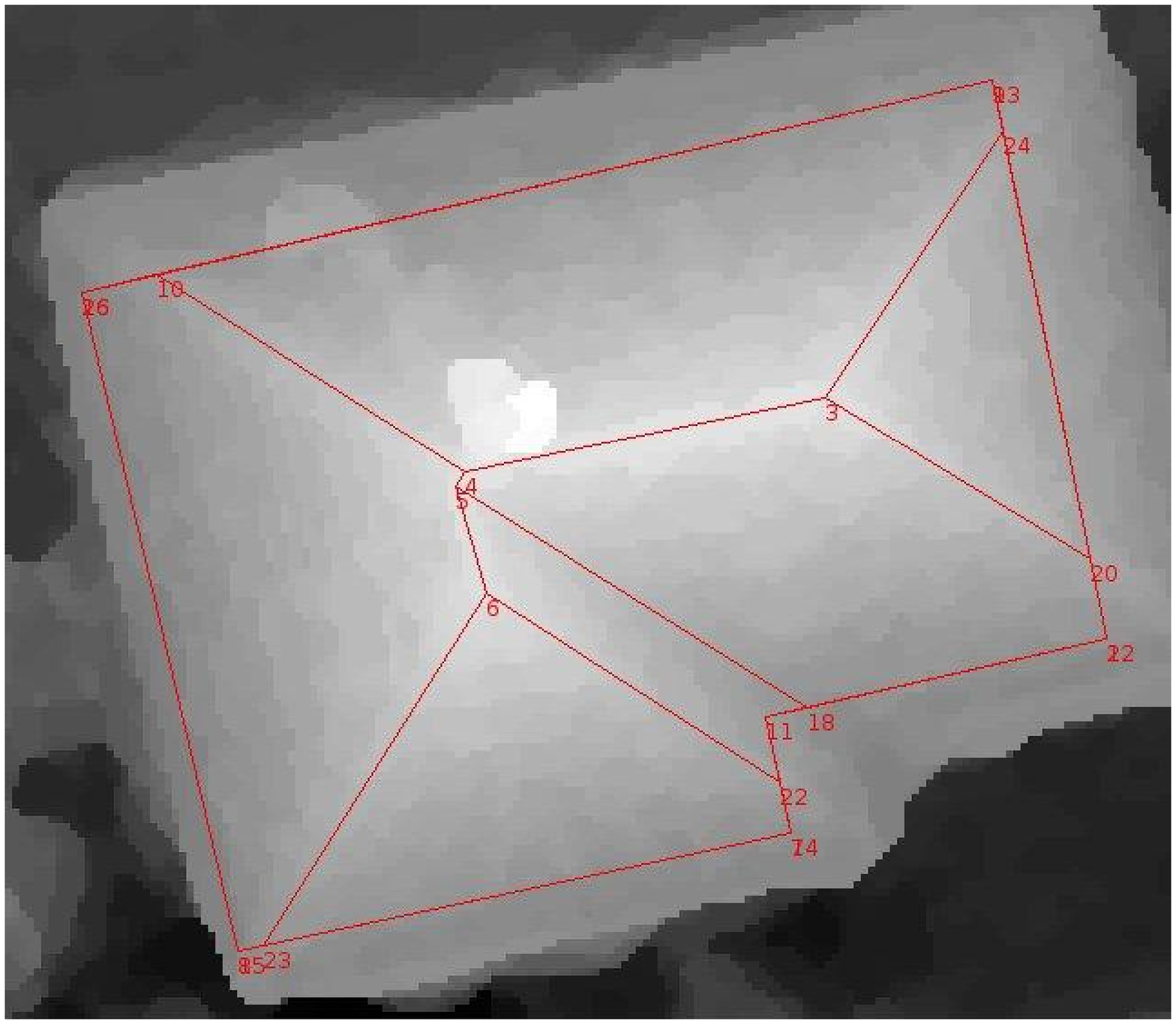}\centering\includegraphics[width=4cm]{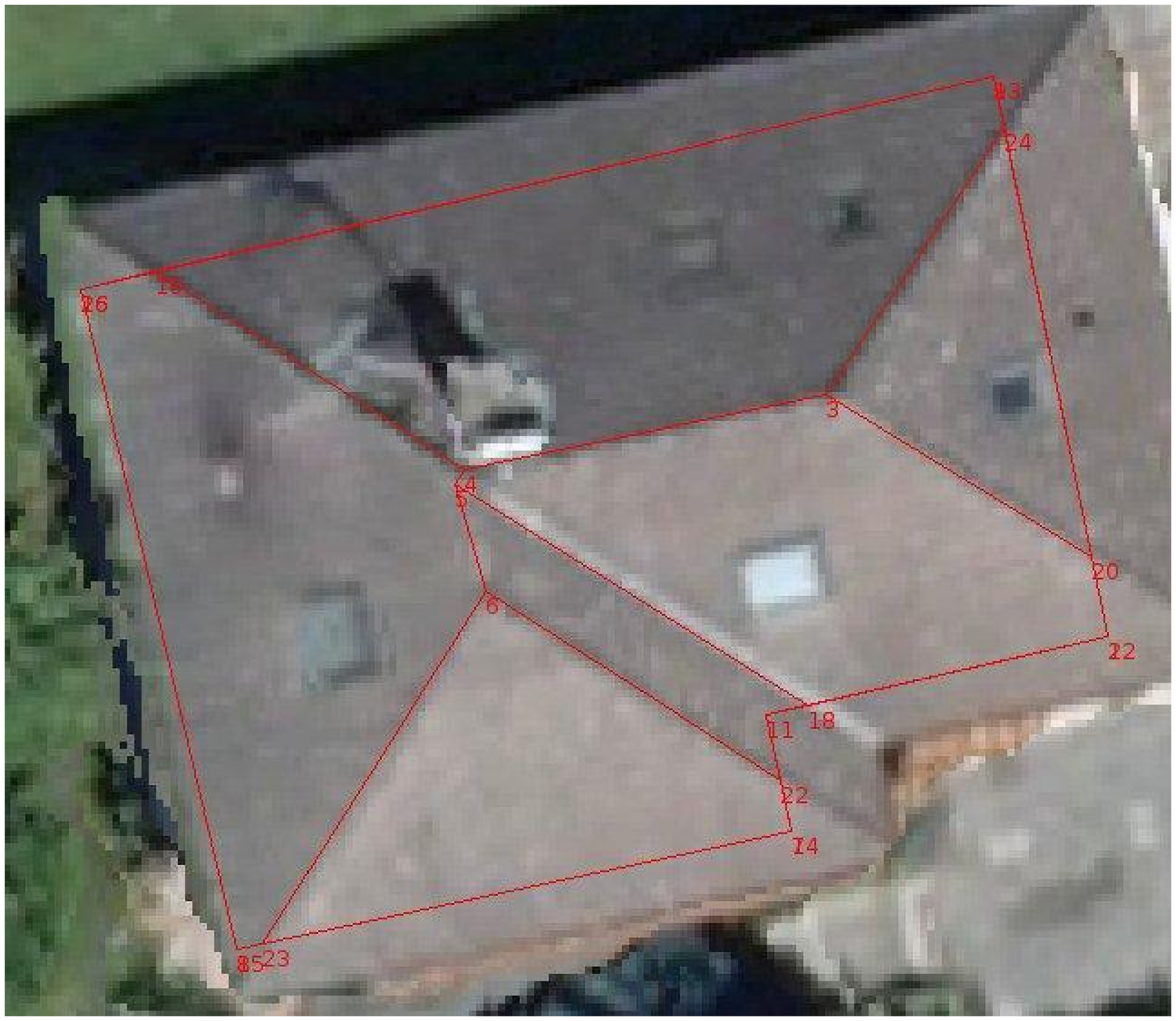}
 \caption{Left background: the DEM. Right background: an orthophotography. First row: the initial building wireframe. Second row: the building at convergence.\label{mb3_Lmo}}
\end{figure}

\begin{figure}
\centering\includegraphics[width=4cm]{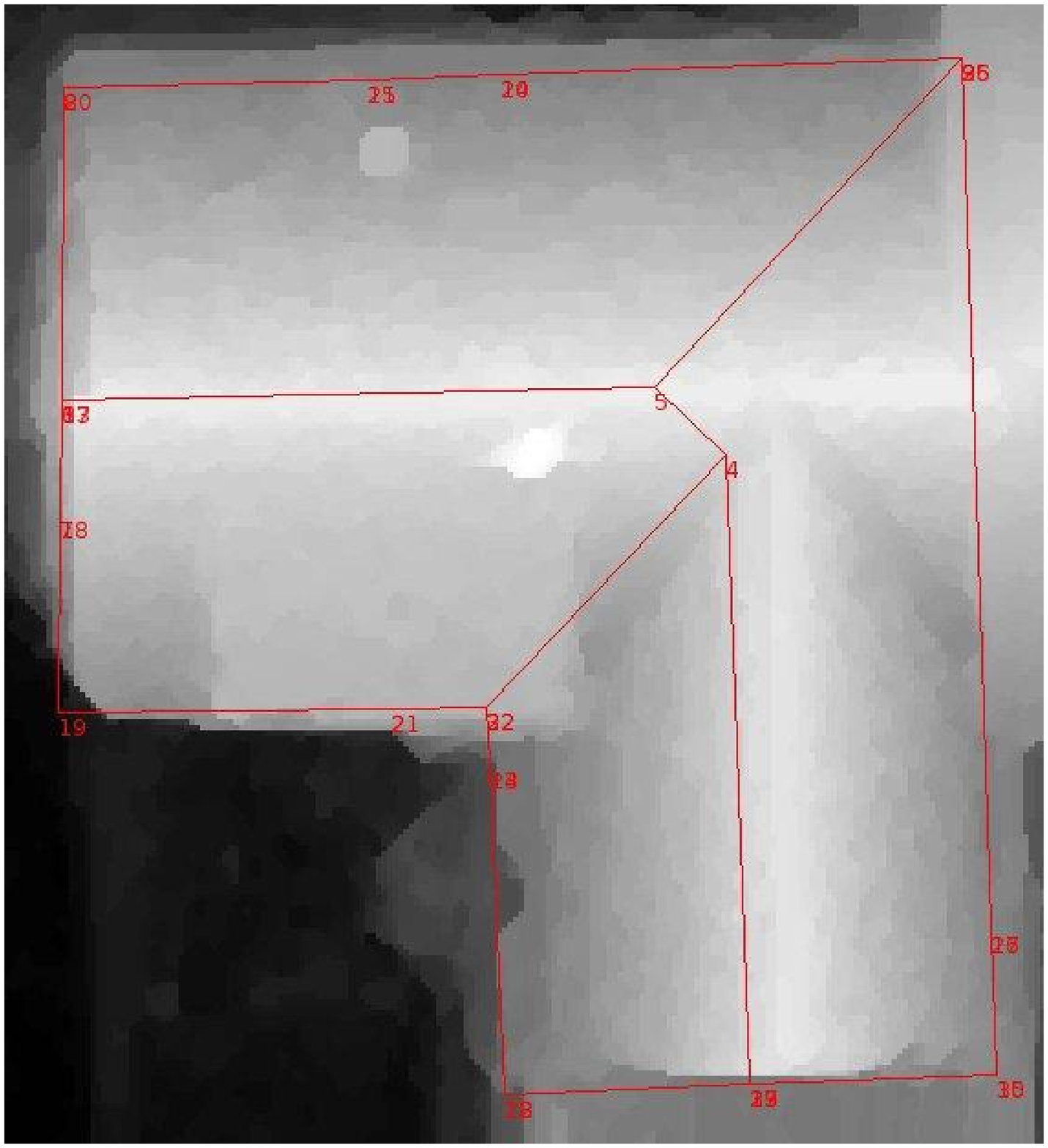}\centering\includegraphics[width=4cm]{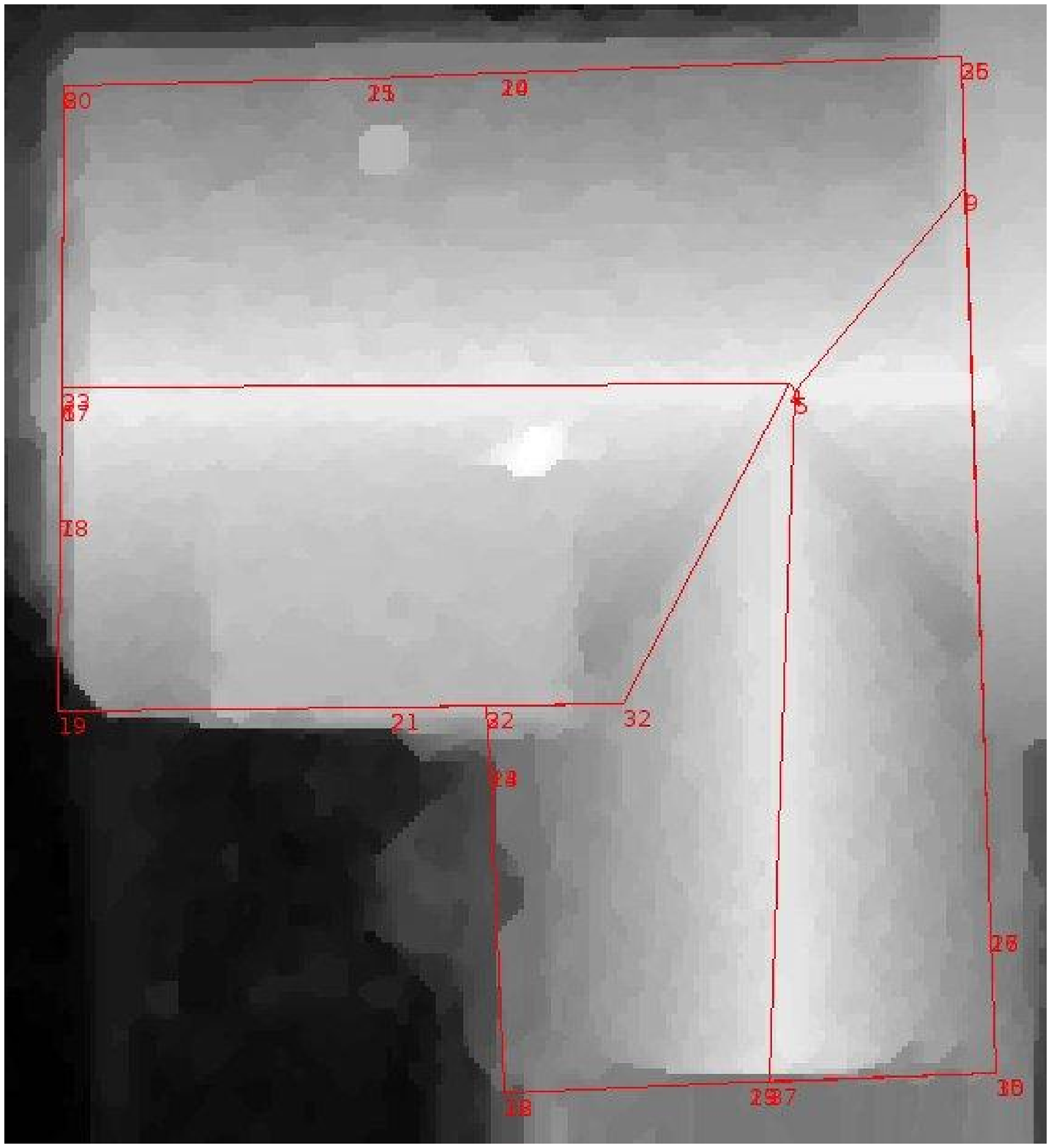}
 \caption{The right part of the roof plane that faces down is missing and cannot be created by our algorithm.\label{mb3_L}}
\end{figure}

\begin{figure}
 \centering\includegraphics[width=4cm]{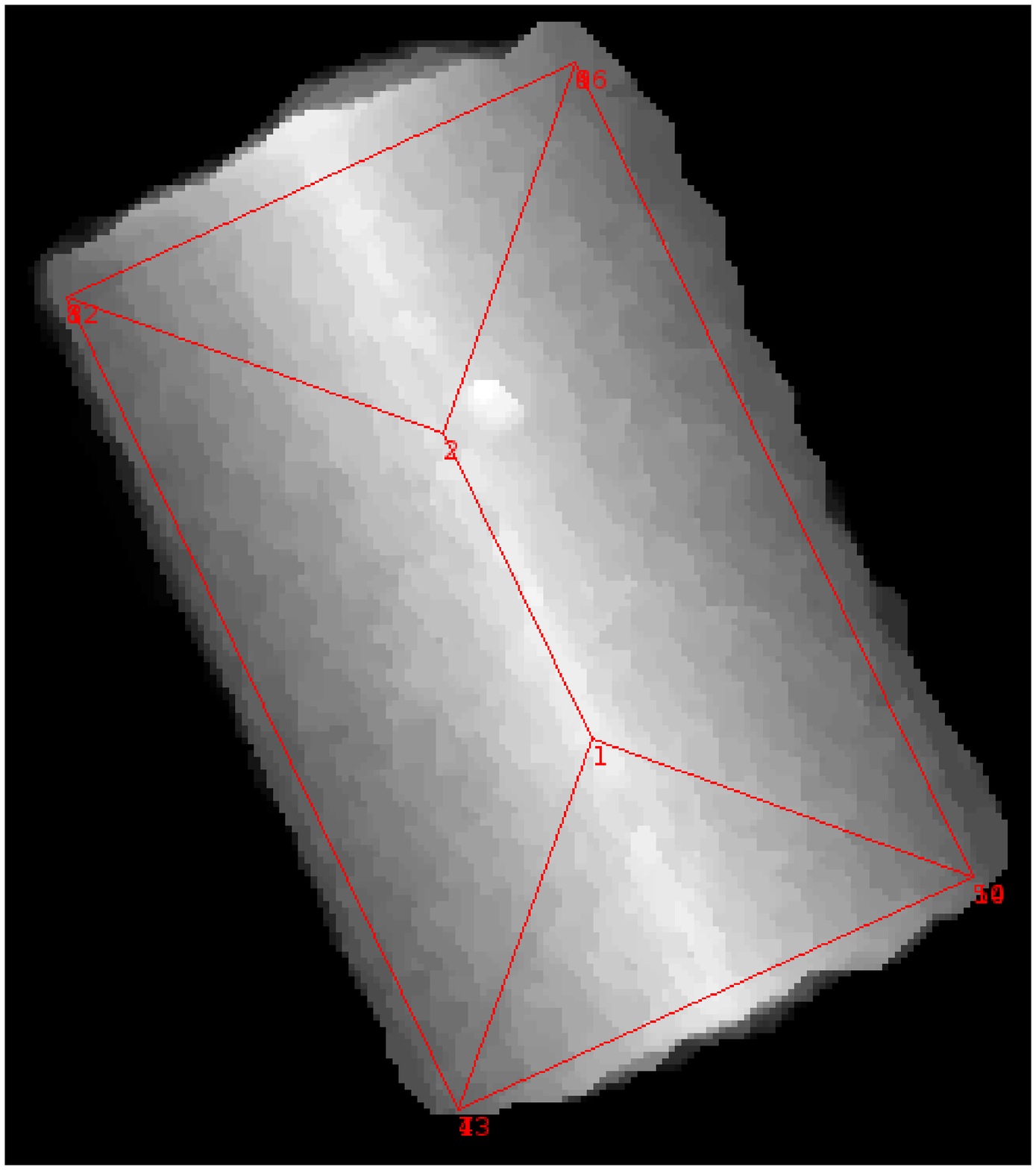}\centering\includegraphics[width=4cm]{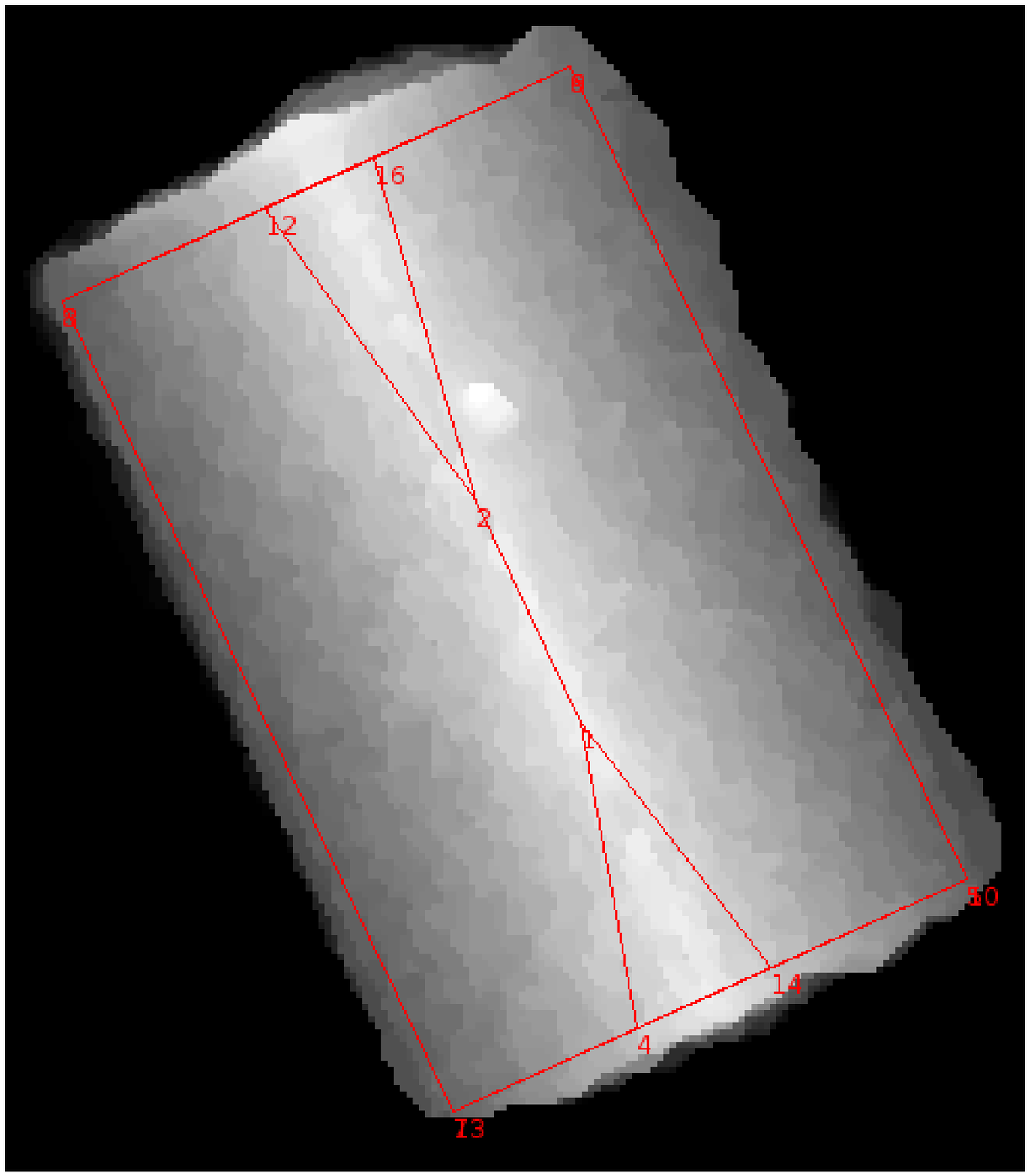}
 \centering\includegraphics[width=4cm]{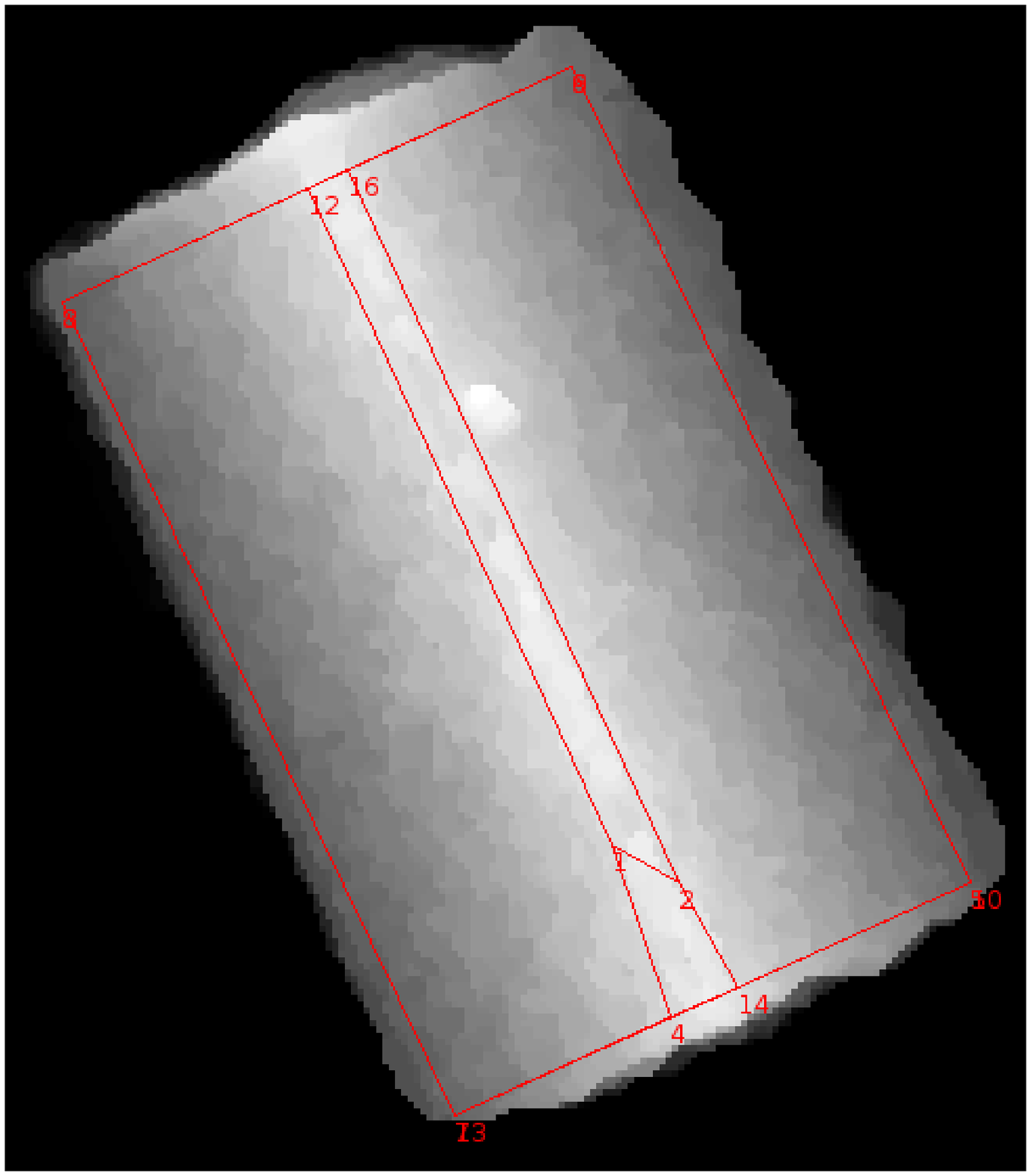}\centering\includegraphics[width=4cm]{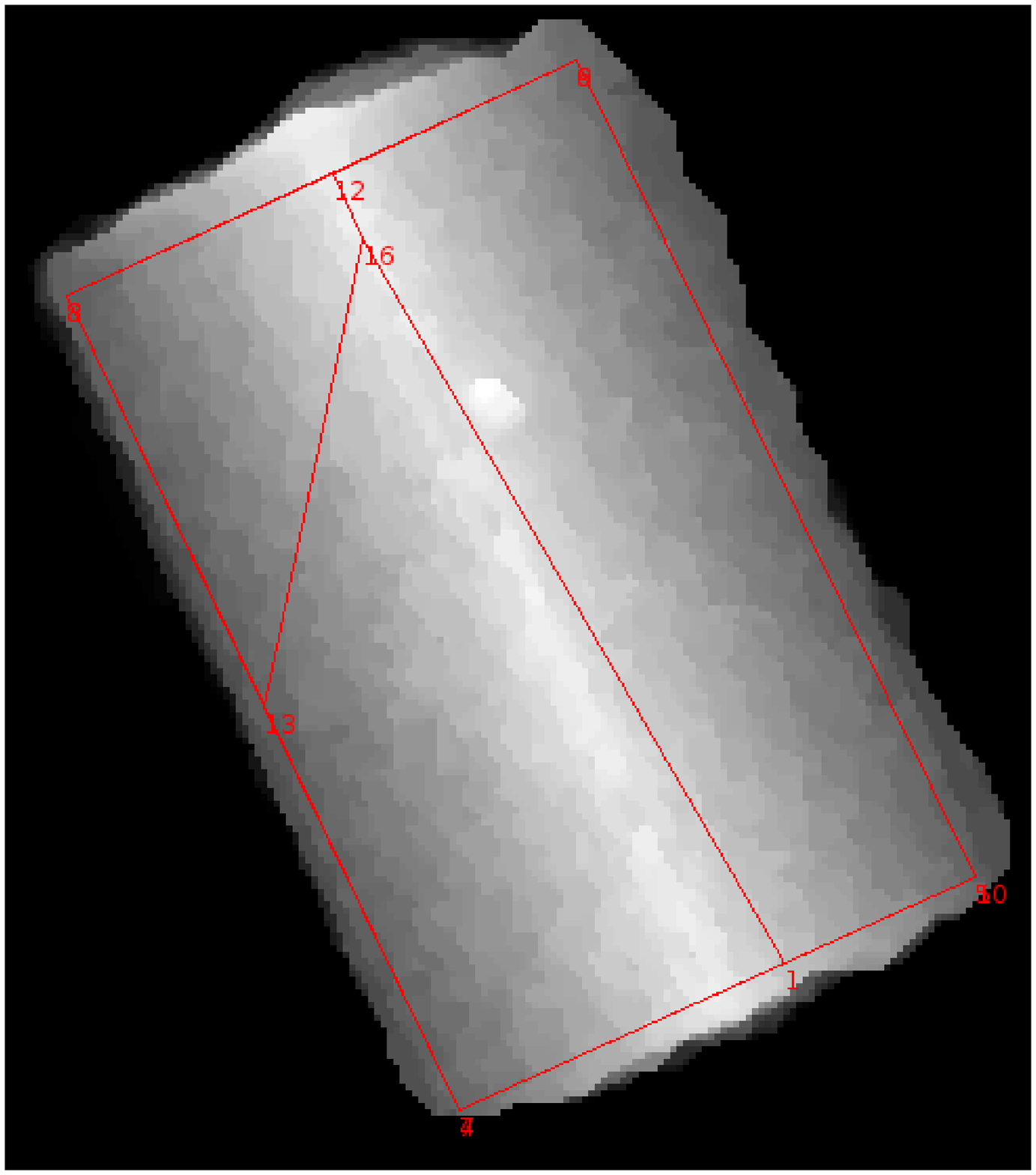}
\centering\includegraphics[width=4cm]{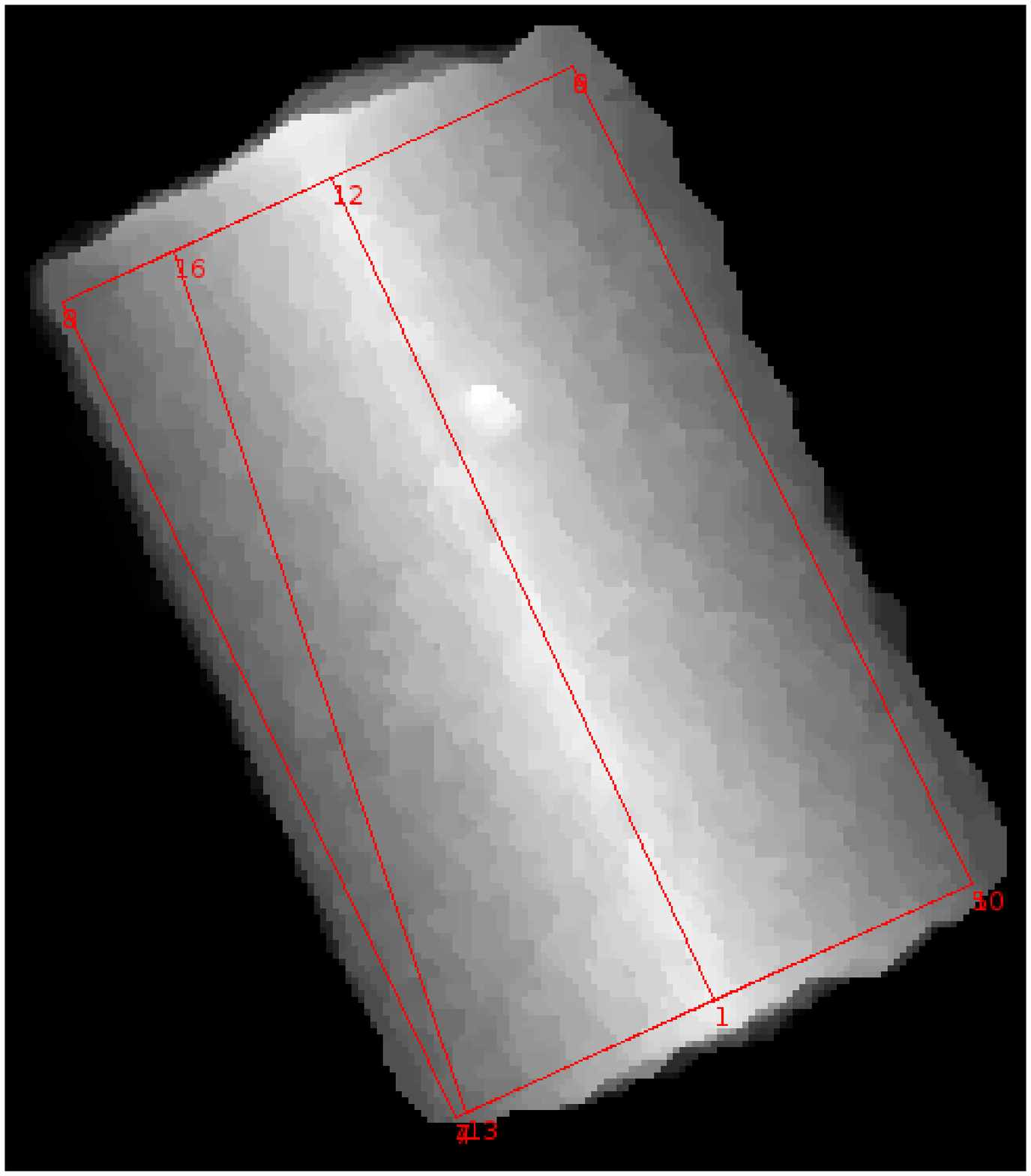}\centering\includegraphics[width=4cm]{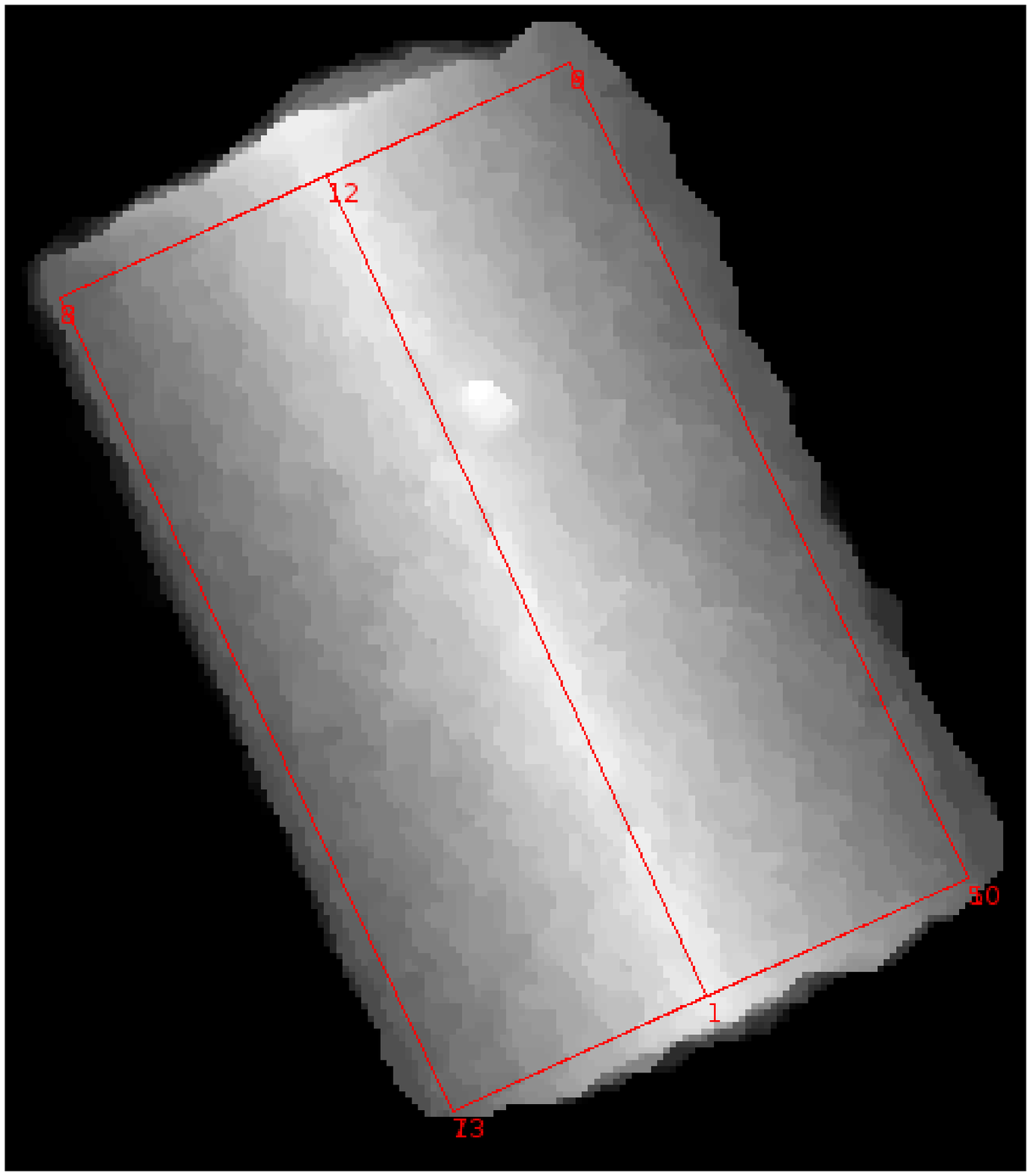}
 \caption{A more complex example were 2 faces are pushed out of the polyhedron.\label{mb3_attaque_v_remove}}
\end{figure}

The typical computing time for the results of the figures \ref{mb3_Lmo},\ref{mb3_hardedgecollapse},\ref{mb3_L},\ref{mb3_attaque_v_remove} is 2 seconds.

Since our algorithm performs minimal topological changes, it never creates any face, so a case like in figure~\ref{mb3_L}, the missing face at the right is not reconstructed. Another possible extension is to check if the faces with high residual errors can be advantageously splitted into some small number of faces, with robustly estimated target plane coordinates.

 In figure \ref{mb3_attaque_v_remove}, the top, bottom and left faces are competing to fit the same data: the points of the DEM of the left roof plane. At an iteration one of the 3 planes estimated to be under the 2 other planes, which resulted in the discarding of the 2 other facets. Another possible outcome could have been that they settle to be distinct but nearly coplanar. Some possible extension could be to detect those nearly coplanar facets to merge them together.

\subsection{Other Applications}

We believe that this framework can have multiple alternative applications. It can be used to perform variational shape approximation\cite{vsa}, but it can also lead to intuitive geometry editing by letting a user move freely planes instead of points or simple face extrusions.

A special case of its application can compute the weighted straight skeleton of a polygon defined in\cite{eppstein99raising,straightskeleton}, when the initial polyhedron faces are the polygon in the plane $z=0$, facing down, another copy, facing up, at $z=1$, and one vertical quadrangular face linking corresponding edges of the 2 horizontal polygons. By decreasing simultaneously the slopes of the vertical faces at a rate given by their weight, until the $z=1$ plane does no longer support any face of the polyhedron, the vertical projection of the resulting polyhedron finally outputs the straight skeleton. As exposed in \cite{eppstein99raising}, the straight skeleton can readily be used in 3D building modelization: the straight skeleton of a floor plan gives the roof topology when there is one roof plane per facade and the slopes are all equal. The weighted straight skeleton possibly computed with our framework even allows the slopes to be different and even negative.

Likewise, a special application of this new framework is to compute straight skeletons of non-convex polyhedra. The plane equation mapping involved is only a translation of the planes at equal velocity without rotations. This is simply the shrinking mapping : $[\vec n,d]\rightarrow [\vec n,d+t]$ if $\vec n$ is normalized. When the evolution has reached the time where the polyhedron is reduced to a single vertex, the loci of the edges spanned during the evolution describe the faces of a straight skeleton of the polyhedron. Recent work\cite{Barequet} has developed a similar kinetic approach specialized to this particular application.

\section{Conclusion}

Our contribution is two-fold: we have introduced a new kinetic framework to allow the direct manipulation of the plane equations of a 3D polyhedron. This framework makes implicitly the minimal changes to the topology of the polyhedron for it to remain bounded and simple. Then we have successfully applied this framework to the 3D building model fitting problem.

Further applicative development will be to use this framework with image based and constraint enforcing fitting\cite{TV05} to relax the previous constraint on the fixed topology, in order to achieve a higher accuracy 3D model.
This research is a building block of a 3D building model enhancement system that simultaneously fit the main planes of a building model and reconstruct its superstructures with parametric objects\cite{BBPM07}.

\bibliographystyle{elsart-num}
\bibliography{bib}

\end{document}